\documentclass[11pt,letterpaper]{article}
\usepackage[utf8]{inputenc}
\usepackage[T1]{fontenc}
\usepackage{babel}
\usepackage{amsmath}
\usepackage{graphicx}
\usepackage{setspace}
\usepackage{microtype}
\usepackage[
  font=doublespacing,
  labelsep=period,
  justification=justified,
  singlelinecheck=false
]{caption}
\usepackage{fancyhdr}
\usepackage{float}

\usepackage[dvipsnames,table]{xcolor}
\usepackage[colorlinks]{hyperref}
\usepackage{xurl} %
\usepackage{multirow}
\usepackage{threeparttable}
\usepackage{enumitem}

\fancyheadoffset[L,R]{\dimexpr(\paperwidth-\textwidth)/2\relax}
\fancyfootoffset[L,R]{\dimexpr(\paperwidth-\textwidth)/2\relax}
\usepackage[margin={1.25in, 1in}]{geometry}
\renewenvironment{abstract}
{%
  \begin{center}
    \bfseries \abstractname
  \end{center}
  \ignorespaces
}
{%
  \par
}

\newlength{\subfiglabelsep}
\newcommand{\subfigimg}[3][,]{%
  \setbox1=\hbox{\includegraphics[#1]{#3}}%
  \begin{minipage}[t]{\wd1}%
    \noindent\textsf{#2}\par
    \vspace{\subfiglabelsep}%
    \usebox1%
  \end{minipage}%
}

\begin{document}

\title{
Harnessing X-ray Absorption Spectroscopy Data through Multimodal Mining of Battery Literature
}

\author{\normalsize
Tanjin He\textsuperscript{1{*}}, 
Aikaterini Vriza\textsuperscript{2}, 
Logan Ward\textsuperscript{1,3}, 
Xu Huang\textsuperscript{4}, \\
\normalsize
Yiming Chen\textsuperscript{2}, 
Anubhav Jain\textsuperscript{5}, 
Gerbrand Ceder\textsuperscript{4}, \\
\normalsize
Rajeev S. Assary\textsuperscript{6}, 
Ian T. Foster\textsuperscript{1,7{*}}, 
Maria K. Y. Chan\textsuperscript{2{*}}}

\date{\vspace{-3ex}}

\maketitle
\thispagestyle{fancy}

{\small
\noindent\textsuperscript{1} Data Science and Learning Division, Argonne National Laboratory, Lemont, IL 60439, USA 

\noindent\textsuperscript{2} Center for Nanoscale Materials, Argonne National Laboratory, Lemont, IL 60439, USA 

\noindent\textsuperscript{3} NVIDIA, Santa Clara, CA 95051, USA 

\noindent\textsuperscript{4} Department of Materials Science and Engineering, University of California, Berkeley, CA 94720, USA 

\noindent\textsuperscript{5} Energy Technologies Area, Lawrence Berkeley National Laboratory, Berkeley, CA 94720, USA 

\noindent\textsuperscript{6} Materials Science Division, Argonne National Laboratory, Lemont, IL, 60439, USA 

\noindent\textsuperscript{7} Department of Computer Science, The University of Chicago, Chicago, IL, 60637, USA

\noindent{*} Corresponding authors: 
Tanjin He (the@anl.gov), 
Ian T. Foster (foster@anl.gov),
Maria K. Y. Chan (mchan@anl.gov)\par

}

\newpage
\begin{abstract}
X-ray absorption spectroscopy (XAS) is central to understanding the local electronic and atomic structure of materials, yet most published spectra remain inaccessible to data-driven analysis because they are embedded in figures and described through fragmented textual context in the literature. 
Here, we use multimodal (image and text) literature mining to transform this dispersed knowledge into an AI-ready experimental data resource.
We developed a scalable spectroscopy data digitization pipeline that identifies XAS figures in full-text articles, digitizes spectral curves, and links each spectrum to accompanying metadata on the measured edge and material. 
Applying this pipeline to the battery literature produced an open dataset of 13,740 XAS spectra, spanning 66 absorbing elements and diverse battery chemistries, with expert validation confirming accurate extraction of spectral and metadata information. 
By converting literature-embedded spectra into structured numerical data, this dataset provides a foundation for large-\allowbreak scale XAS analysis, cross-\allowbreak laboratory comparison, high-\allowbreak throughput characterization, and autonomous discovery of advanced materials.
\end{abstract}

\newpage
\section*{Background \& Summary}

The explosive growth of AI/ML in materials science necessitates high-quality, large-scale, labeled, experimental and computational datasets for training. 
While there are ample computational data, experimental data availability is far more limited. 
The vast amount of data embedded in the scientific literature represents an important, largely untapped source for enriching experimental datasets.

Large-scale spectroscopy datasets can power materials characterization, autonomous discovery, and cross-laboratory standardization.
First, they provide references for identifying structural and compositional information in complex materials, extending existing databases \cite{xaslib,ishii2025GlobalCrossd,spasyuk2025XASDBDesign,paripsa2024RefXASOpena,kieffer2016SSHADEFAME} toward broader coverage of materials and experimental conditions.
Second, they supply the training data needed to develop the high-throughput characterization algorithms that enable self-driving laboratories \cite{chen2024RobustMachin,jia2026RevealLocal,liu2025AutomaDeterm, fei2026agentic, huang2025cascade, huang2025skillpuzzler,penfold2024machine}.
Finally, by spanning many research groups, they can help establish standardized experimental procedures and reporting ontologies for cross-laboratory alignment.

Scientific publications have accumulated a large amount of spectroscopy data that offer a natural complementary source for growing curated reference databases.
However, these data are typically presented as scientific figures with textual descriptions scattered throughout the paper, and this unstructured, multimodal format hinders their use in data-driven research.
Wang et al.~\cite{wang2025x} compiled a valuable dataset of 1,652 X-ray absorption spectroscopy (XAS) spectra of iron-containing proteins from the literature, but the manual curation involved is labor-intensive and limits scalability.
Automated literature mining is becoming more tractable with advances in natural language processing (NLP) and computer vision (CV), and especially with the growing capabilities of large language models (LLMs) and vision language models (VLMs).
For example, He et al.~\cite{he2023PrecurRecomm} and Huo et al.~\cite{huo2022MachinRation} extracted materials synthesis recipes from the literature for predictive synthesis. 
In a related vein, Leong et al.~\cite{leong2025MERMaiUniver,leong2024AutomaElectr} extracted chemical reaction information for organic synthesis and photocatalysis. 
Beyond reaction data, Zheng et al.~\cite{zheng2024ImageDataa} mined the reticular chemistry literature.
Dong et al.~\cite{dong2022auto} extracted semiconductor band-gap data from text.
Liu et al.~\cite{liu2026MultimDatase} extracted causal mechanisms for materials design and performance optimization. 
Siegel et al.~\cite{siegel2018extracting} trained a neural network to detect bounding boxes of scientific figures.
Park et al.~\cite{park2021advances} finetuned an embedding model to find explanatory sentences most similar to XAS figure captions.
More recently, Zhang et al.~\cite{zhang2026DIVEHydrog} targeted multimodal information on hydrogen storage materials.

The most challenging part of this multimodal data extraction task is extracting data from figures, where most characterization and performance data reside \cite{sayeed2024NLPMeets}.
Current VLMs can reliably interpret object-level information in figures, such as categorizing an image \cite{zheng2024ImageDataa}, recognizing text within a figure \cite{leong2025MERMaiUniver,leong2024AutomaElectr}, or generating a qualitative description of the image \cite{liu2026MultimDatase}, but they still struggle to capture pixel-level information such as precise curve coordinates, which matter most because they represent the actual spectroscopy data.
For example, MatGD \cite{lee2024MatGDMateri} reported only 66\% accuracy for data-line separation, with particular difficulty when data lines share similar colors or overlap significantly.
Similarly, DIVE \cite{zhang2026DIVEHydrog} found that reading key points from data curves with a VLM (Gemini-2.5-Flash + Deepseek-R1) required up to a 50\% relative-error tolerance because of visual reading noise.
Circi et al.~\cite{circi2025InformExtrac} likewise reported that the precision of o1 in extracting tabular data from figures ranged from only 23\% to 45\%.
Typical errors included hallucinated points at convenient x-values, trend-like but imprecise coordinates, and missed tightly clustered or overlapping points.
Our own tests directly applying VLMs (GPT-5.2 and Opus 4.6) to extract data from XAS figures show similar issues (Supplementary Figs.~S1--S4).
Together, these studies highlight substantial room for improvement in multimodal data extraction.
These limitations indicate that dedicated models are needed to complement current VLMs and to form a pipeline that balances precise and flexible extraction.
Leong et al.~\cite{leong2025MERMaiUniver}, for instance, demonstrated the advantages of combining a chemical-structure-recognition model, RxnScribe \cite{qian2023RxnScrSequen}, with a VLM to extract multimodal reaction information.
Spectroscopy data extraction, however, remains underexplored, and it is the focus of this work.

\begin{figure}[H]
	\centering
	\includegraphics[width=1.0\linewidth]{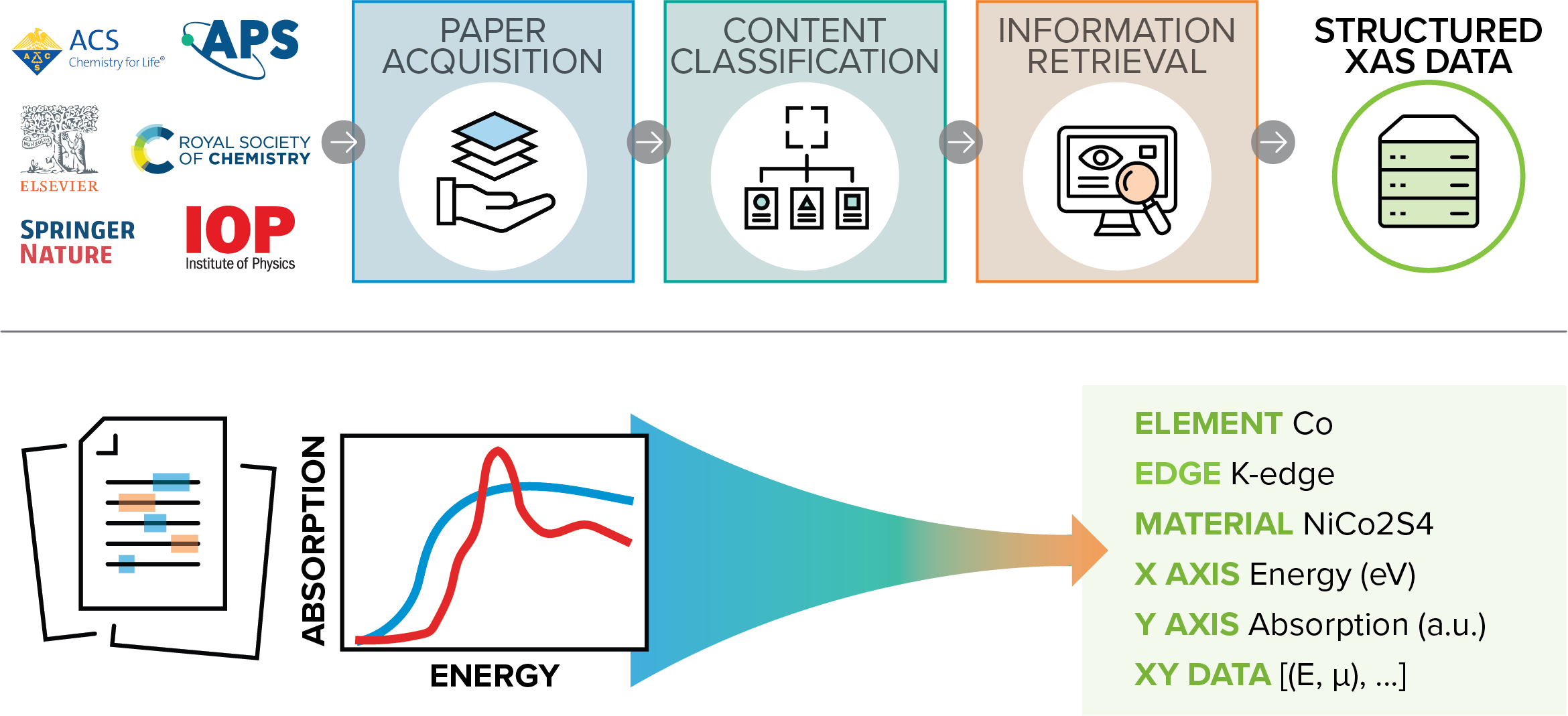}
	\caption{
		Schematic of the spectroscopy data digitization pipeline proposed in this work. 
		Top panel: the pipeline converts publisher articles into a structured XAS dataset via paper acquisition, content classification, and information retrieval. 
		Bottom panel: example of processing an XAS figure and its contextual text into a machine-readable record containing the absorbing element, absorption edge, material, axis labels, and energy-dependent absorption coordinates.
	}
	\label{fig:workflow}
\end{figure}

In this work, we provide a fully auto-generated, open-source dataset of 13,740 XAS spectra retrieved from 3,510 battery-related papers.
We target XAS because it is a powerful element-specific probe of oxidation state and local atomic structure, yet costly to measure.
Building on our previous work, including EXSCLAIM \cite{schwenker2023EXSCLAHarnes}, Plot2Spectra \cite{jiang2022Plot2SAutoma}, and figure extraction for optical emissivity \cite{baibakova2022OpticaEmissi}, we developed a scalable spectroscopy data digitization pipeline (Fig.~\ref{fig:workflow}) that searches for and downloads full-text papers, classifies images by spectroscopy relevance, as well as orchestrates multimodal agents for curve separation, axis and legend recognition, and contextual information extraction.
Starting from 485,628 battery-related papers and 4,112,327 figures, the pipeline filtered the collection down to XAS-relevant figures and successfully digitized 13,740 XAS curves into machine-readable numerical spectral data, each accompanied by metadata on the measured edge and material.
The dataset is publicly available in JSONL format, joining the community effort to grow curated databases and make more XAS data findable and reusable.
By making these data readily accessible, the digitized dataset lowers the barrier to data-driven analysis of XAS measurements and facilitates the characterization of materials and electrochemical systems for autonomous discovery.

\section*{Methods}

\subsection*{Paper acquisition}

To keep the scope manageable, we focused on battery-related papers in this work rather than the entire body of scientific publications, as batteries constitute a field in which advanced characterization techniques in general, and XAS in particular, are widely applied to diverse materials.
The publications used in this work are journal articles from Springer Nature, Elsevier, the Royal Society of Chemistry (RSC), IOP Publishing (IOP), and the American Physical Society (APS), for which we obtained text and data mining permissions, together with open-access papers from the American Chemical Society (ACS).
Using EXSCLAIM to leverage each publisher's search engine, we identified papers containing the keywords ``battery'' or ``batteries'' and retrieved their Digital Object Identifiers (DOIs).
We then used Playwright \cite{playwright} and the publishers' APIs to download the full-text articles in HTML/XML format. 
High-resolution images were also downloaded as needed for downstream processing. 
Because full-text articles published before 1995 are mostly in PDF format, which complicates data processing, we restricted our search to papers published from 1995 onward.
All data are stored and managed in an internal document-oriented database implemented in MongoDB.
Finally, the HTML/XML papers were parsed into natural-language paragraphs and figure captions using LimeSoup \cite{kononova2019TextmiDatase}, which removes irrelevant markup while preserving the paper structure and section headings.

\subsection*{Content classification}

Figures in scientific papers are highly flexible and vary widely in content.
In this work, we focus on XAS figures for data extraction because XAS measurements are costly, typically requiring competitively allocated synchrotron beamtime.
Its element specificity and sensitivity to local electronic and atomic structure make it especially well suited for tracking oxidation-state and coordination changes in battery research. 
Recovering XAS data from the published literature is therefore particularly valuable. 
To identify XAS figures of interest, we use a multi-stage filtering process in which inexpensive filters are applied first, followed by more sophisticated filters once the dataset has been reduced to a more manageable size.

The filtering proceeds in three stages.
We first classify figures based on their captions using a self-hosted open-source LLM, Mixtral-8x22B-Instruct-v0.1, to determine whether each figure contains spectroscopy results and, specifically, whether it corresponds to XAS. 
We then download only the XAS figures to conserve download bandwidth.
Because the downloaded figures may be either single-panel figures or multi-panel composites of several subfigures, we use FigureSeparator \cite{jiang2021TwostaFramew} in EXSCLAIM to separate multi-panel figures into individual subfigures.
Finally, we apply a VLM, GPT-5.2 (gpt-5.2-2025-12-11), to classify each subfigure according to whether it shows typical X-ray absorption near edge structure (XANES) spectra, presented as line plots of absorption intensity as a function of photon energy.
Subfigures showing non-XAS data, Fourier-transformed (FT) XAS data (FT magnitude versus radial distance), extended X-ray absorption fine structure (EXAFS) oscillations (signal versus wavevector in $k$-space), or in-situ XAS visualizations (waterfall or colormap plots) are excluded.

\subsection*{Information retrieval}

For each XAS figure, our information-retrieval workflow consists of multiple agents that analyze the figure, caption, and paper text to extract the XAS results and the necessary metadata for identifying the XAS measurement, including the axis information, precise curve coordinates, curve legend, absorbing element, absorption edge, measured material and constituent elements.
More details are provided in the following subsections. 
GPT-5.2 is used as the VLM model in the information-retrieval workflow.

\subsubsection*{Axis understanding}
\label{sec:axis-understanding}

Axis labels and tick information are extracted using a combination of optical character recognition (OCR) and a VLM.
Because the VLM is adept at capturing text and understanding its role in a figure, we first use it to determine which text corresponds to the axis labels, i.e., the text identifying the physical quantity and unit of each axis.
The VLM does not, however, provide accurate text positions; we therefore employ an OCR model, PP-OCRv5~\cite{cui2025Paddle30}, to capture the values and positions of the axis tick labels, i.e., the numerical values marked along each axis.
We retain only those tick labels whose values are cross-checked by the VLM.
The resulting tick label positions and values are used to construct a linear transformation that converts the pixel positions of curve data points into energy-dependent absorption-intensity coordinates.

\subsubsection*{Curve segmentation}
\label{sec:curve-segmentation}

To capture the precise curve coordinates of XAS results, we developed a two-step curve segmentation algorithm based on connectivity and color decomposition.
Because a typical XAS figure contains multiple curves that overlap, intersect, and span the full width of the plot, we use color signals to identify the color-consistent portions of a target curve and connectivity information to recover the remaining parts, which are connected to these portions but may be occluded by other curves.
While this algorithm mitigates confusion from elements disconnected from the curves, such as annotations and legends, it cannot recover dashed or dotted lines, which we leave for future work.

We first convert the plot into a line-segment connectivity graph, in which each segment is a connected, non-background pixel region between two adjacent intersection points of the curves.
The curve segmentation task then reduces to finding the combination of line segments that best reconstructs a single complete curve trace corresponding to one XAS measurement.
To identify this combination, we apply the BIRCH algorithm~\cite{zhang1996birch} to decompose the plot into clusters of similarly colored pixels, and we assign higher reward to combinations with higher color consistency, i.e., those whose pixels predominantly belong to a single cluster.
We further introduce a roughness penalty that suppresses the sharp transitions arising when segments from different curves are joined, thereby favoring combinations that trace a smooth path.
The reward function is defined in Eq.~\ref{eq:reward}:
\begin{equation}
\mathrm{reward}(\ell) = \frac{\left| \ell \cap C_{\text{color}} \right|}{W \, \hat{w}} - \frac{\lambda}{N} \sum_{i=1}^{N} \left| f''(x_i) \right|,
\label{eq:reward}
\end{equation}
where $\ell$ denotes a candidate line (combination of segments), $C_{\text{color}}$ is the set of pixels in a color-decomposition cluster, $W$ is the image width, $\hat{w}$ is the estimated line width of $\ell$, $f$ is the centerline trajectory of the finite-width plotted curve $\ell$, $f''$ is its second derivative, $\{x_i\}_{i=1}^{N}$ are the $N$ points along $f$, and $\lambda$ controls the roughness penalty.
The value of $\lambda$ was chosen empirically (as 0.01) so that the roughness penalty takes effect only when the color-consistency term alone cannot distinguish between candidate lines. 
With the curves separated, we convert each centerline trajectory from pixel positions to energy-dependent absorption-intensity coordinates using the linear transformation constructed from the axis tick label positions and values in Section~{\hypersetup{linkcolor=blue}\nameref{sec:axis-understanding}}.

\subsubsection*{Legend recognition}

Based on the curves separated in Section~{\hypersetup{linkcolor=blue}\nameref{sec:curve-segmentation}}, we attribute the legends in the figure to their corresponding curves.
For each separated curve, we synthesize a side-by-side highlighting visualization that places the original figure on the left and an inpainted version on the right, in which the target curve is emphasized while the other visual components are faded.
We then ask GPT-5.2 to interpret this highlighting visualization and recognize the legend corresponding to the emphasized curve.
In practice, an ensemble approach that queries the VLM multiple times improves the stability and accuracy of inference. 
We therefore repeat the procedure twice using two slightly different inpainting strategies: in one, the highlighted curve is rendered with the average color of its pixels, while in the other it retains its original pixel-level color variation.
We accept a legend assignment only when the two inferences agree, thereby trading recall for precision.

\subsubsection*{Contextual metadata extraction}

To contextualize each XAS measurement, we extract the necessary metadata from the figure caption and paper text, ensuring that the resulting data record is self-contained and interpretable without reference to the original paper.

Figure legends often contain only the minimal information needed to distinguish curves within a single figure, such as sample identifiers or the values of key experimental variables; additional context is therefore required to link each legend entry to the corresponding explanations in the caption and the detailed descriptions in the main text.
Because the full paper text is typically lengthy and is reused for extracting different types of metadata, we first use GPT-5.2 to condense it by extracting verbatim descriptions relevant to the figure, including discussions of the figure itself; explanations of abbreviations, sample identifiers, and terminology; and descriptions of the experimental conditions associated with the measurement.
We then provide GPT-5.2 with the highlighting visualization of the target curve, its associated legend, the figure caption, and the extracted figure-related descriptions, and ask it to identify the absorbing element, the absorption edge, and the material being measured together with its constituent elements.

\subsection*{Dataset generation}

Using the spectroscopy data digitization pipeline, we searched and found 485,628 battery-related papers and downloaded full text for 460,440 of them.
The downloaded papers contain 4,112,327 figure captions. 
After content classification, 6,055 figures were identified as figures of interest that contain line plots of XAS spectra and were downloaded for downstream processing.
Through information retrieval, 4,312 of the 6,055 figures yielded at least one complete curve-level record, resulting in 13,740 digitized XAS curves with machine-readable numerical spectral data and accompanying metadata on the measured edge and material.

\section*{Data Records}

We provide the complete dataset of 13,740 XAS spectra as a single JSONL file, which will be available at Figshare upon publication.
Each record corresponds to a single XAS curve, together with metadata describing its source and measurement details.
The detailed data schema is shown in Table~\ref{tab:tab1}, and a complete example record is provided in Supplementary Fig.~S5.
The visual components extracted from each figure are stored in the
\texttt{xy\_data},
\texttt{x\_\allowbreak axis\_\allowbreak label},
\texttt{y\_\allowbreak axis\_\allowbreak label},
and \texttt{legend} fields.
The spectral data points are stored in \texttt{xy\_data} as a list of [x, y] pairs, where x is the photon energy and y is the absorption intensity, as indicated by the x- and y-axis labels.
The \texttt{x\_\allowbreak axis\_\allowbreak label} and \texttt{y\_\allowbreak axis\_\allowbreak label} fields preserve the original axis text from the figure. 
The former contains the energy unit, typically eV and occasionally keV. 
The latter either labels the absorption intensity explicitly as ``a.u.'' or leaves it unspecified, as the intensity is given in arbitrary units and only its relative magnitude is meaningful.
The \texttt{legend} field stores the original text of the figure legend corresponding to the spectrum, serving as an anchor that links the spectrum to the figure caption and the paper text for further details.

In addition to the visual components, metadata describing each spectrum are stored in the \texttt{measured\_\allowbreak element}, \texttt{measured\_\allowbreak edge}, \texttt{material}, and \texttt{elements} fields.
The \texttt{measured\_\allowbreak element} field contains the chemical symbol of the absorbing element, and \texttt{measured\_\allowbreak edge} specifies the absorption edge of the spectrum, typically denoted K, L$_2$, L$_3$, etc.
The \texttt{material} field reports the name or chemical formula of the measured material, drawn from the figure legend and supplemented with contextual information from the paper. 
For example, if the legend gives only a sample code, this field provides a concise chemical definition of the code together with the relevant experimental conditions.
The \texttt{elements} field lists the chemical symbols of the constituent elements of the material, enabling element-based queries across the dataset.

To support source traceability and access to additional details when needed, the dataset provides the \texttt{doi} and \texttt{figure\_\allowbreak label} fields. 
The DOI identifies the original paper, and the figure label locates the corresponding figure within that paper.
The original figures and paper text are not included in the dataset, in accordance with text and data mining agreements.

\begin{table}[H]
\centering
\small
\begin{tabular}{|p{2.7in}|l|l|}
\hline
\rowcolor{YellowGreen}
\textbf{Data description} & \textbf{Data Key Label} & \textbf{Data Type} \\
\hline
DOI of the source paper & \texttt{doi} & \emph{string} \\
\hline
Figure label within the source paper & \texttt{figure\_label} & \emph{string} \\
\hline
Legend of spectrum in the source figure & \texttt{legend} & \emph{string} \\
\hline
$x$-axis label of the spectrum & \texttt{x\_axis\_label} & \emph{string} \\
\hline
$y$-axis label of the spectrum & \texttt{y\_axis\_label} & \emph{string} \\
\hline
Spectrum data points, shape $(N,2)$ & \texttt{xy\_data} & \emph{list} of [\emph{float}, \emph{float}] \\
\hline
Absorbing element & \texttt{measured\_element} & \emph{string} \\
\hline
Absorption edge (e.g.\ K, L$_2$) & \texttt{measured\_edge} & \emph{string} \\
\hline
Material name or chemical formula & \texttt{material} & \emph{string} \\
\hline
Constituent elements of the material & \texttt{elements} & \emph{list} of \emph{strings} \\
\hline
\end{tabular}
\caption{
Schema of each dataset record, with fields for source traceability (\texttt{doi}, \texttt{figure\_label}), figure-\allowbreak derived spectral content (\texttt{legend}, \texttt{x\_axis\_label}, \texttt{y\_axis\_label}, \texttt{xy\_data}), and contextual measurement metadata (\texttt{measured\_element}, \texttt{measured\_edge}, \texttt{material}, \texttt{elements}).
}
\label{tab:tab1}
\end{table}

\section*{Technical Validation}

We validate the dataset through manual annotation to quantify extraction accuracy and through statistical analyses to characterize its diversity and coverage.

\subsection*{Extraction accuracy}

To evaluate the accuracy of the extracted data, we randomly sampled 100 records from the dataset for verification by materials science experts (annotation interface in Supplementary Fig.~S6).
For the figure components, the experts visually inspected each figure to confirm that the extracted axis labels, tick labels, and legend entries matched the corresponding text in the original figure.
Tick-label placement was verified by overlaying the extracted tick values on the original figure at their extracted positions and checking their alignment with the corresponding tick labels.
The tick labels of an axis are counted as correct if every extracted label has the correct value and position and at least two are extracted, the minimum needed for the linear transformation to data coordinates (Section~{\hypersetup{linkcolor=blue}\nameref{sec:axis-understanding}}).
Curve separation was assessed by overlaying the extracted curve on the original figure and verifying that it followed the original trace without visible discrepancy.
To calibrate this visual criterion, we manually annotated 10 curves point by point; curves judged to be in visual agreement had deviations averaging <1\% of the intensity range and falling below 3\% for 99\% of points.

For the metadata, the experts checked the measured element, edge, material, and constituent elements against the legend and paper text to confirm consistency with the original figure and accompanying text.
Each record was expected to specify one elemental absorption edge, such as Fe K or Ni L, for the extracted curve.
The \texttt{material} field was expected to resolve the legend into a material name or chemical formula, especially when the legend mentioned only an experimental variable and omitted the chemical identity, which then had to be supplemented from the paper text.
The \texttt{elements} field was evaluated against the material composition, with missing or extra elements counted as errors.

The accuracy results are summarized in Table~\ref{tab:tab2}.
Axis information is extracted with essentially perfect accuracy, reaching 100\% for both labels and ticks.
Legend recognition reaches 97\% accuracy. 
This task is more challenging than simply reading text from a figure, because the VLM must distinguish individual curves and establish a one-to-one correspondence between data curves and legend entries. 
Legend errors occur primarily when neighboring curves are visually similar or overlap heavily, leading the model to assign an incorrect legend entry to a curve.
Curve separation achieves 91\% accuracy and is the most challenging task. 
Because our manual evaluation requires each extraction to capture the entire curve, even a single local error is counted as an invalid extraction. 
In the failure cases, the discrepancy is typically localized to a small region where multiple curves intersect or overlap extensively, while the majority of the curve remains well captured (examples in Supplementary Fig.~S7).
Metadata extraction is nearly perfect, reflecting the proficiency of GPT-5.2 on textual tasks. 
Accuracy is 100\% for both the measured element and the edge, 97\% for the material, and 98\% for its constituent elements.
Errors in the material field are correlated with legend recognition, since an incorrect legend can lead to an incorrect material name or a misspecification of the experimental conditions. 
Constituent-element errors are similarly linked to material-field errors, occurring when the chemical portion of the extracted material is incorrect.
Taken together, 89\% of the sampled records have all figure components and metadata extracted correctly.

\begin{table}[ht]
\centering
\small
\renewcommand{\arraystretch}{1.5}
\begin{tabular}{|l|c||l|c|}
\hline
\rowcolor{YellowGreen}
\multicolumn{2}{|c||}{\textbf{(a) Figure components}} &
\multicolumn{2}{c|}{\textbf{(b) Metadata}} \\
\hline
\rowcolor{YellowGreen!40}
\textbf{Attribute} & \textbf{Accuracy} (\%) & \textbf{Attribute} & \textbf{Accuracy} (\%) \\
\hline
Axis labels   & 100 & Measured element & 100 \\
Axis ticks       & 100 & Measured edge    & 100 \\
Legend           & 97 & Material         & 97 \\
Curve separation & 91 & Elements         & 98 \\
\hline
\end{tabular}
\caption{
Accuracy of the spectroscopy data digitization pipeline, evaluated on 100 records randomly sampled from the dataset and verified by materials science experts. 
}
\label{tab:tab2}
\end{table}

\subsection*{Dataset mining}

To assess the diversity and coverage of the dataset, we conducted statistical analyses on its temporal, application, compositional, and inter-laboratory aspects.

\begin{figure}[H]
    \centering
    \begin{tabular}{@{}p{0.462\linewidth}@{}p{0.538\linewidth}@{}}
        \subfigimg[width=\linewidth]{\textbf{(a)}}{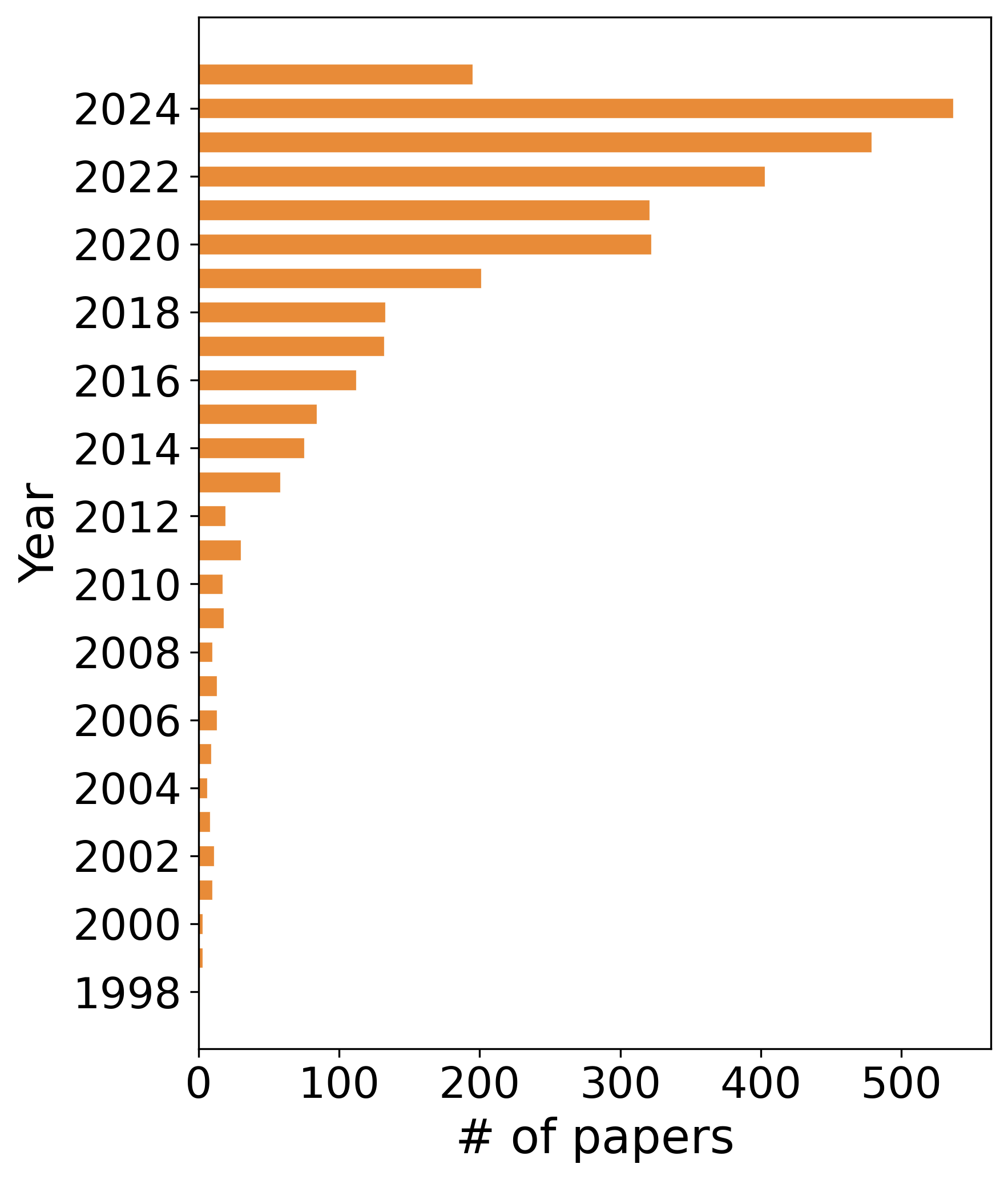} &
        \subfigimg[width=\linewidth]{\textbf{(b)}}{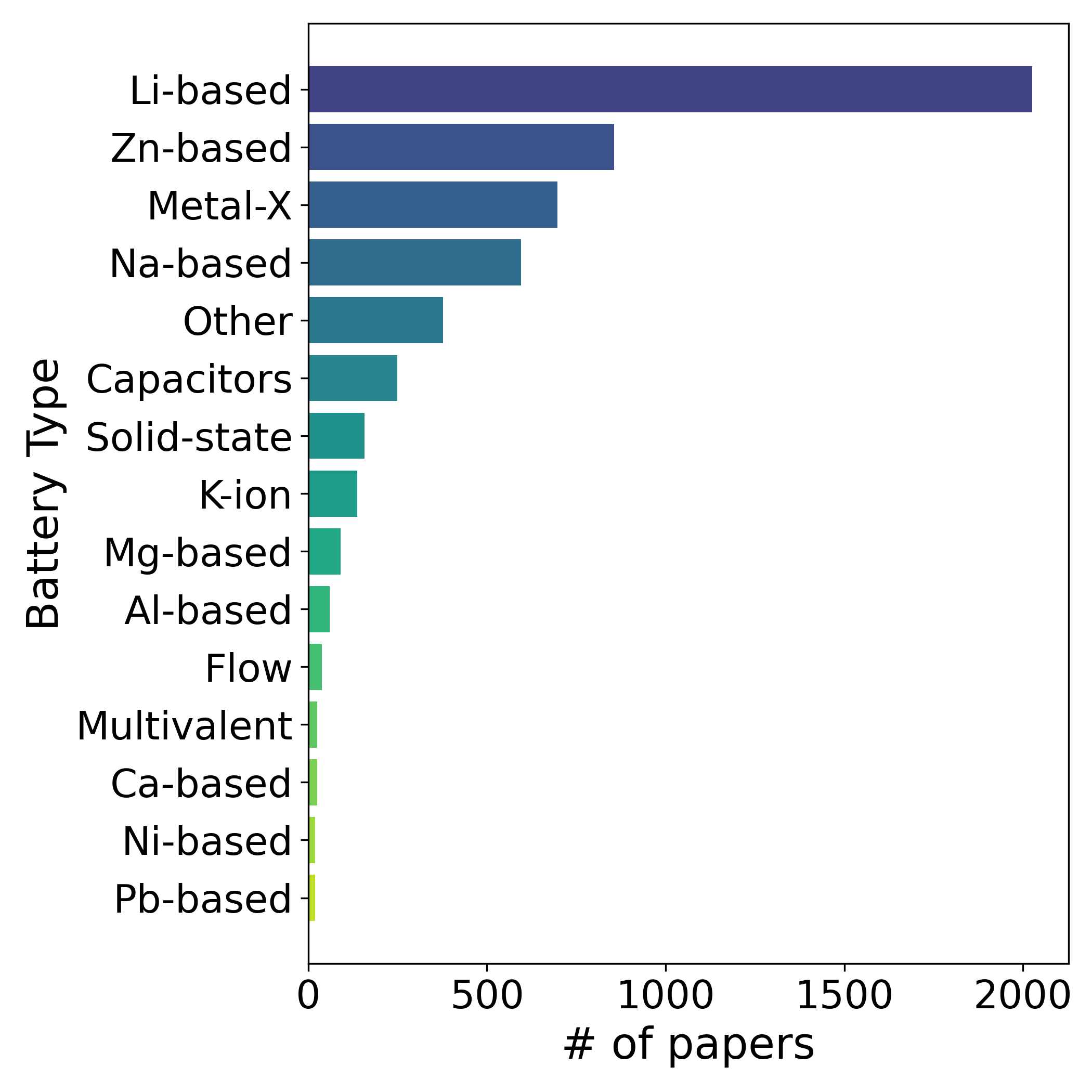}
    \end{tabular}
	\caption{
        Distribution of source papers from which XAS spectra were extracted. 
        \textbf{(a)} Paper counts by publication year.
        \textbf{(b)} Paper counts by battery application. General mentions without a specific metal are grouped as Metal-X (X = air, sulfur, CO$_2$, etc.), Solid-state, and Multivalent.
	}
	\label{fig:paper_stats}
\end{figure}

We first examined the distribution of the source papers by publication year and application background (Fig.~\ref{fig:paper_stats}).
After the digitization pipeline, the 13,740 spectra in the final dataset originate from 3,510 papers.
The distribution by publication year is shown in Fig.~\ref{fig:paper_stats}(a).
The number of source papers grows approximately exponentially over time, consistent with the general trend of scientific publishing.
Because our collection ends in mid-2025, the bar for 2025 is truncated.
The continued growth of relevant papers offers an opportunity to build a larger dataset by reapplying the same pipeline over a longer collection period.

The battery applications mentioned in the introduction sections of the source papers are summarized in Fig.~\ref{fig:paper_stats}(b).
Because a given paper may address more than one application, papers may be counted under multiple battery types.
The source papers span a broad range of battery chemistries---including Li-, Na-, K-, Zn-, and Mg-based systems---along with more general mentions that do not specify a particular chemistry, such as ``solid-state battery'' or ``metal-air battery.''
Because the dataset is drawn from the recent characterization literature, it is weighted toward chemistries under active investigation; mature commercial systems such as lead-acid and nickel–metal hydride (NiMH) are sparsely represented.
This broad coverage demonstrates the effectiveness of the literature-mining approach in capturing diverse domain-specific information, illustrated here by battery applications.
Li-based batteries account for the largest share, consistent with their commercial success and sustained prominence in battery research.
Zn-based batteries also represent a notable portion of this XAS-focused dataset, largely owing to studies of Zn–air batteries, a widely investigated metal–air system in which XAS is routinely used to probe transition-metal species involved in the oxygen reduction and oxygen evolution reactions (ORR/OER)~\cite{chen2025multi, chen2023adsgt}.

\begin{figure}[H]
	\centering
	\includegraphics[width=1.0\linewidth]{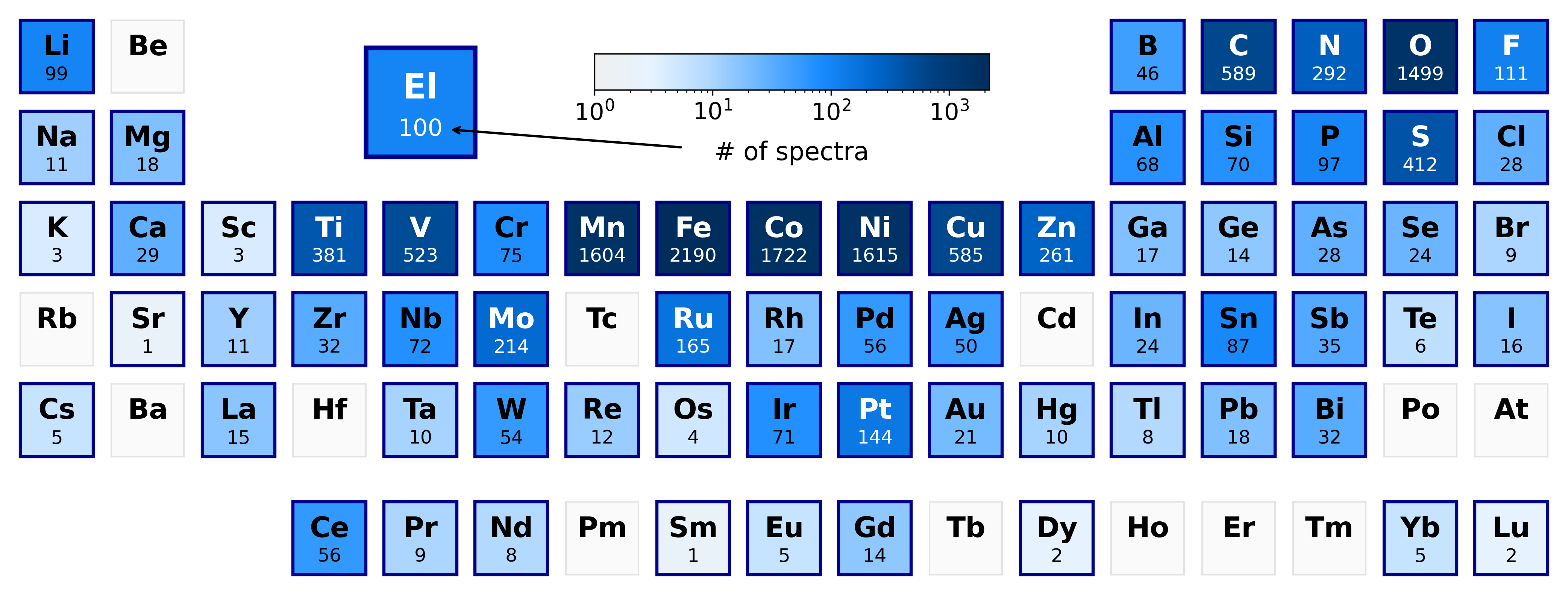}
	\caption{
		Elemental coverage of the dataset, presented as a periodic-table heatmap of the number of XAS spectra measured for each absorbing element.
        Counts are reported within each element box and encoded using a logarithmic white-to-blue color scale.
	}
	\label{fig:ele_table}
\end{figure}

Next, we evaluate the chemical space covered by the dataset.
The number of XAS spectra measured for each element is mapped onto the periodic table in Fig.~\ref{fig:ele_table} using a white-to-blue color gradient, with the count shown in each element box.
The dataset spans a broad chemical space of 66 elements.
The most frequently measured elements are Fe, Co, Ni, and Mn, reflecting the predominance of cathode materials based on LiFePO$_4$ and lithium nickel cobalt manganese oxide (NCM).
Many moderately measured elements are associated with substitution or doping strategies within these frameworks, such as Al substitution and Zr, W, or B doping for enhanced structural stability and cycling performance.
Beyond these conventional frameworks, alternative cathode chemistries such as disordered rock-salt (DRX) materials are represented in the dataset.
DRX materials combine redox-active metals (e.g., Mn, V, Cr) with high-valent $d^0$ stabilizers (e.g., Ti$^{4+}$, Nb$^{5+}$, Mo$^{6+}$), reducing dependence on Co and Ni while achieving high capacity.
The inclusion of these elements makes the dataset a useful reference for resolving XAS spectral diversity arising from multiple oxidation states and varied local environments.
The dataset also spans a wide energy range, from hard X-rays for heavy metals, such as Pb or Bi, to soft X-rays for nonmetal elements, such as C or O.

\begin{figure}[H]
    \centering
    \begin{tabular}{@{}p{0.5\linewidth}@{}p{0.5\linewidth}@{}}
        \subfigimg[width=\linewidth]{\textbf{(a)}}{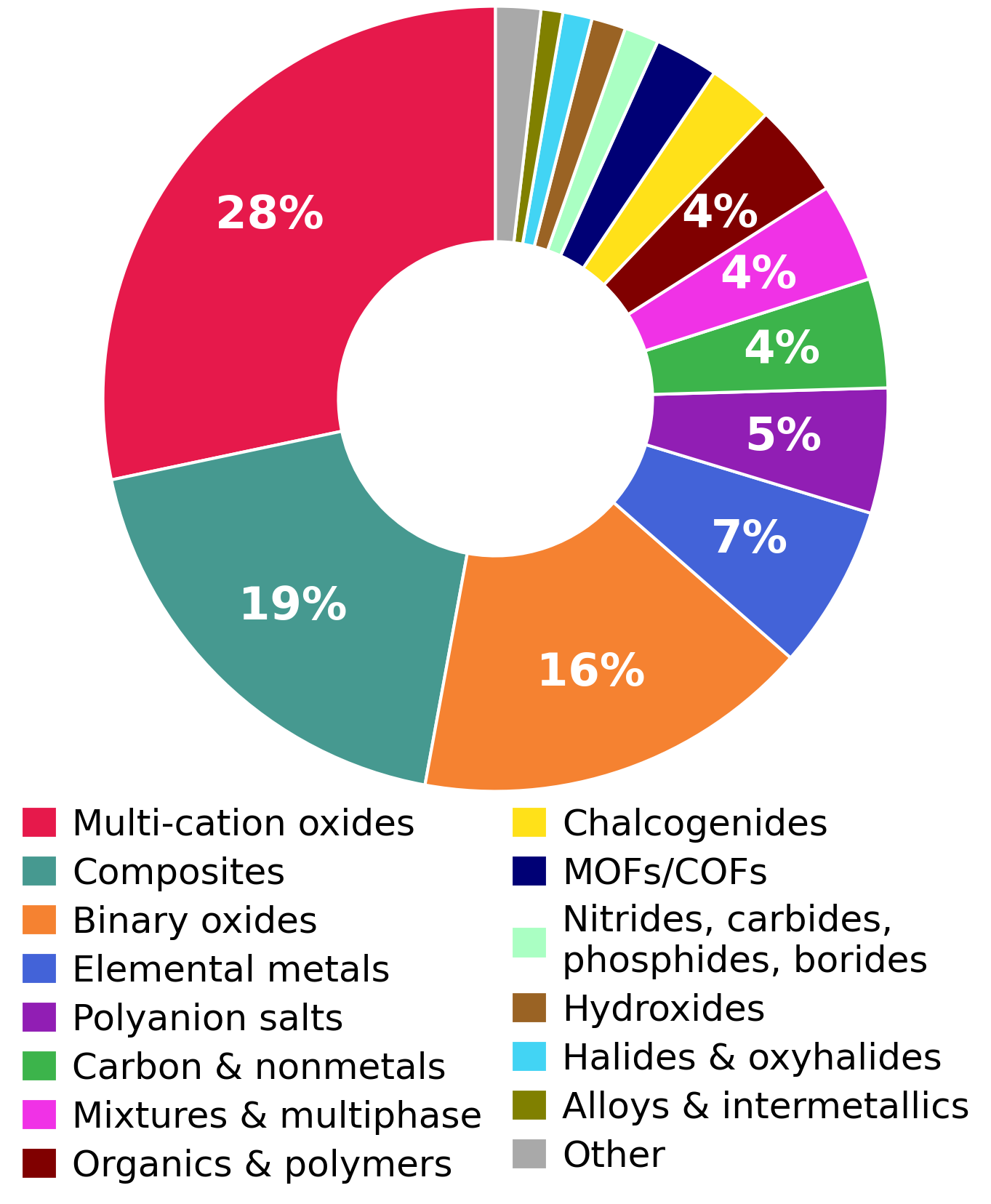} &
        \subfigimg[width=\linewidth]{\textbf{(b)}}{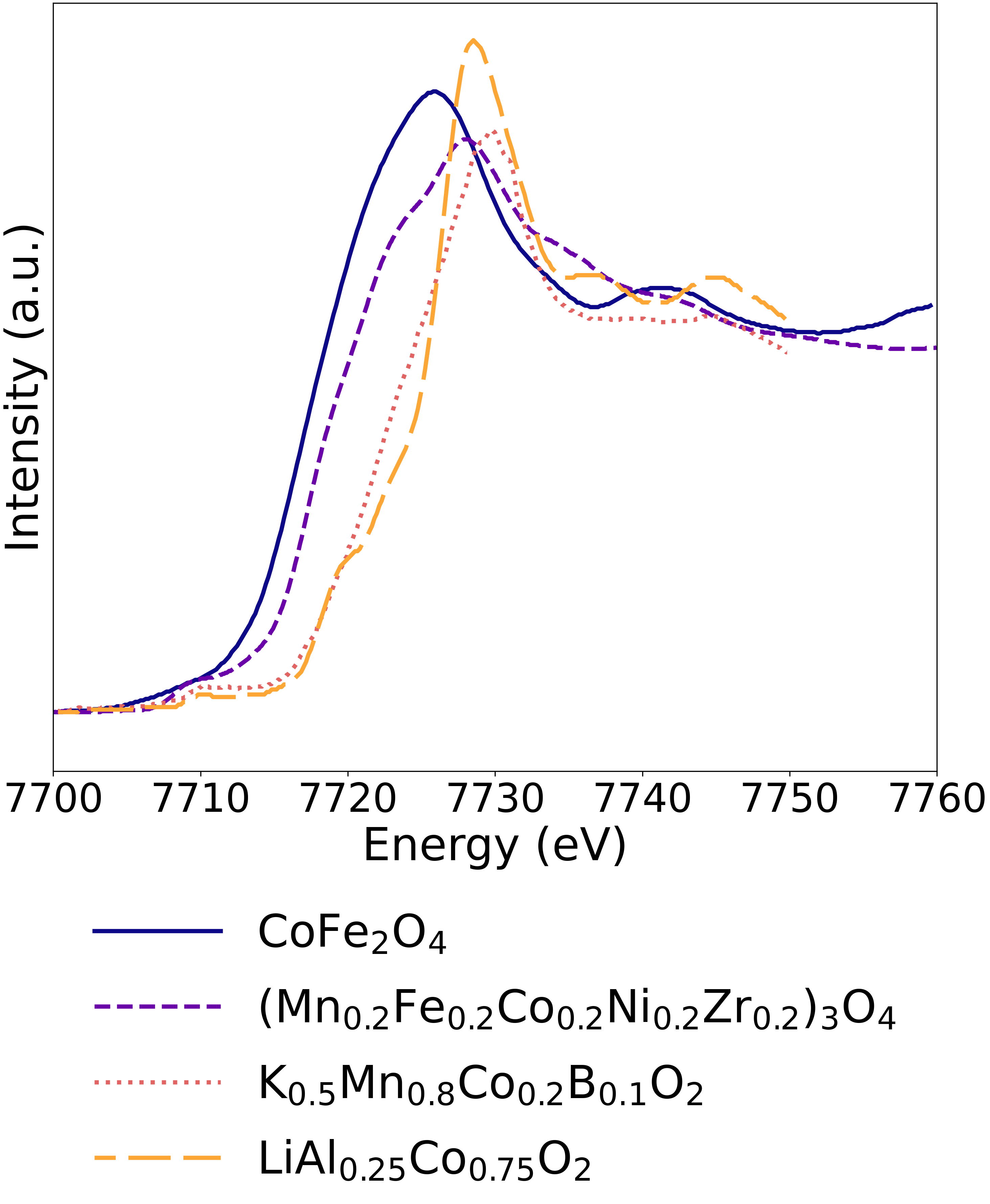}
    \end{tabular}
	\caption{
        Chemical diversity and complexity of the materials in the dataset. 
        \textbf{(a)} Distribution of material classes by composition.
        \textbf{(b)} Representative Co K-edge XAS spectra of multi-component oxide materials beyond binary oxides, including ternary, quaternary, quinary, and high-entropy oxides. 
	}
	\label{fig:mats_distribution}
\end{figure}

We also examine the chemical complexity of the materials in the dataset.
As shown in Fig.~\ref{fig:mats_distribution}(a), the dataset spans a broad range of material classes, including oxides and hydroxides, composites and mixtures, metals and alloys, non-oxide anion compounds, and framework or molecular materials.
Oxides are the most frequently measured class, comprising both binary and multi-cation oxides, reflecting their widespread use as cathodes, solid electrolytes, and certain anodes in batteries.
Composites, which are engineered architectures of two or more materials, also constitute a notable portion of the dataset, consistent with their use as electrocatalysts (e.g., Fe--N--C in Zn--air batteries), heterostructured electrodes (e.g., NCM@rGO), and coated materials (e.g., LiFePO$_4$/C).
Beyond simple materials such as elemental metals and binary oxides, the dataset includes multi-component compounds, such as ternary, quaternary, quinary, and high-entropy oxides (Fig.~\ref{fig:mats_distribution}(b)).
The inclusion of these complex materials provides reference spectra for resolving the local atomic environments of advanced material systems.

\begin{figure}[H]
    \centering
    \begin{minipage}[c]{0.6\linewidth}
        \subfigimg[width=\linewidth]{\textbf{(a)}}{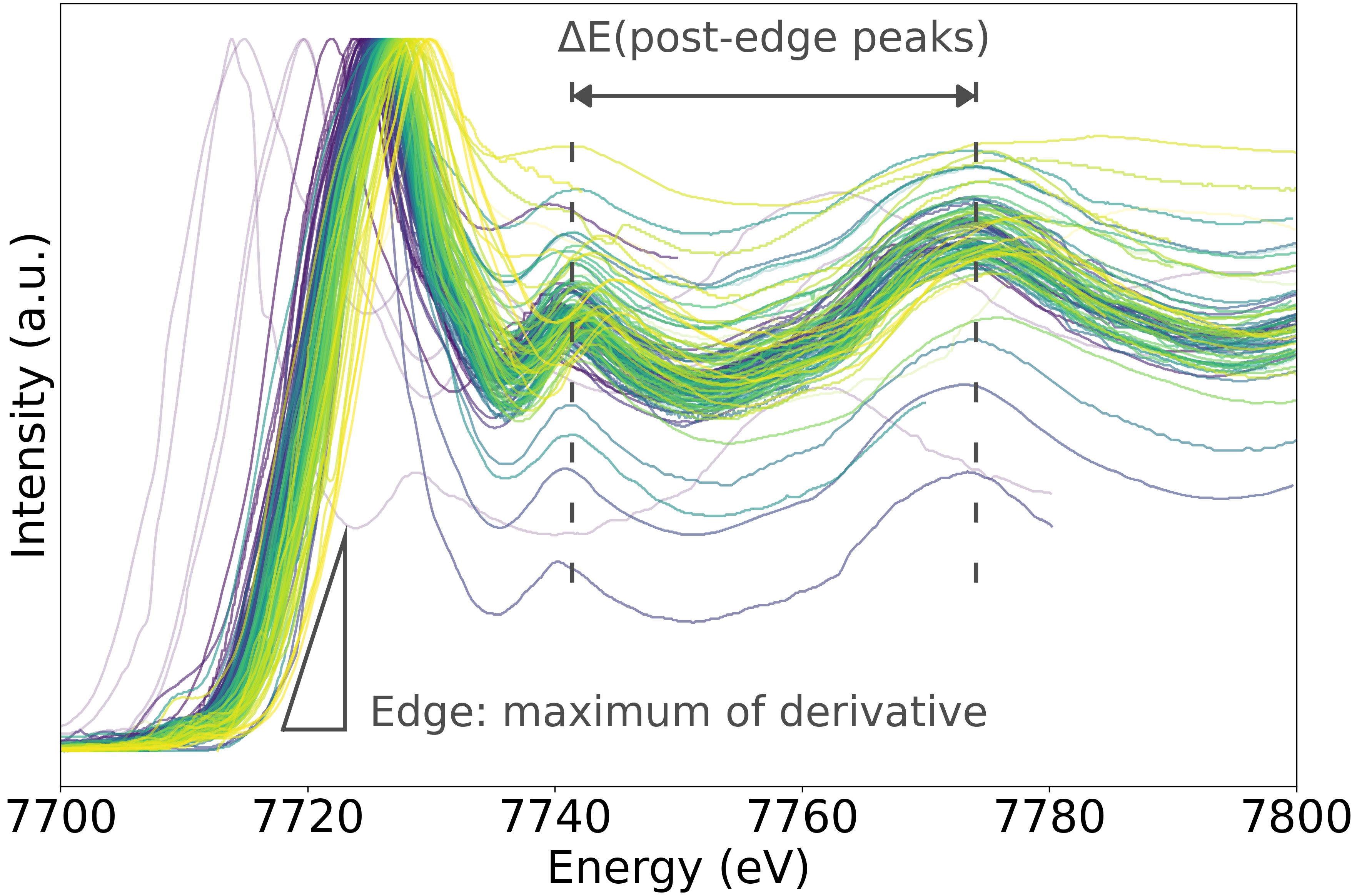}
    \end{minipage}%
    \begin{minipage}[c]{0.4\linewidth}
        \subfigimg[width=\linewidth]{\textbf{(b)}}{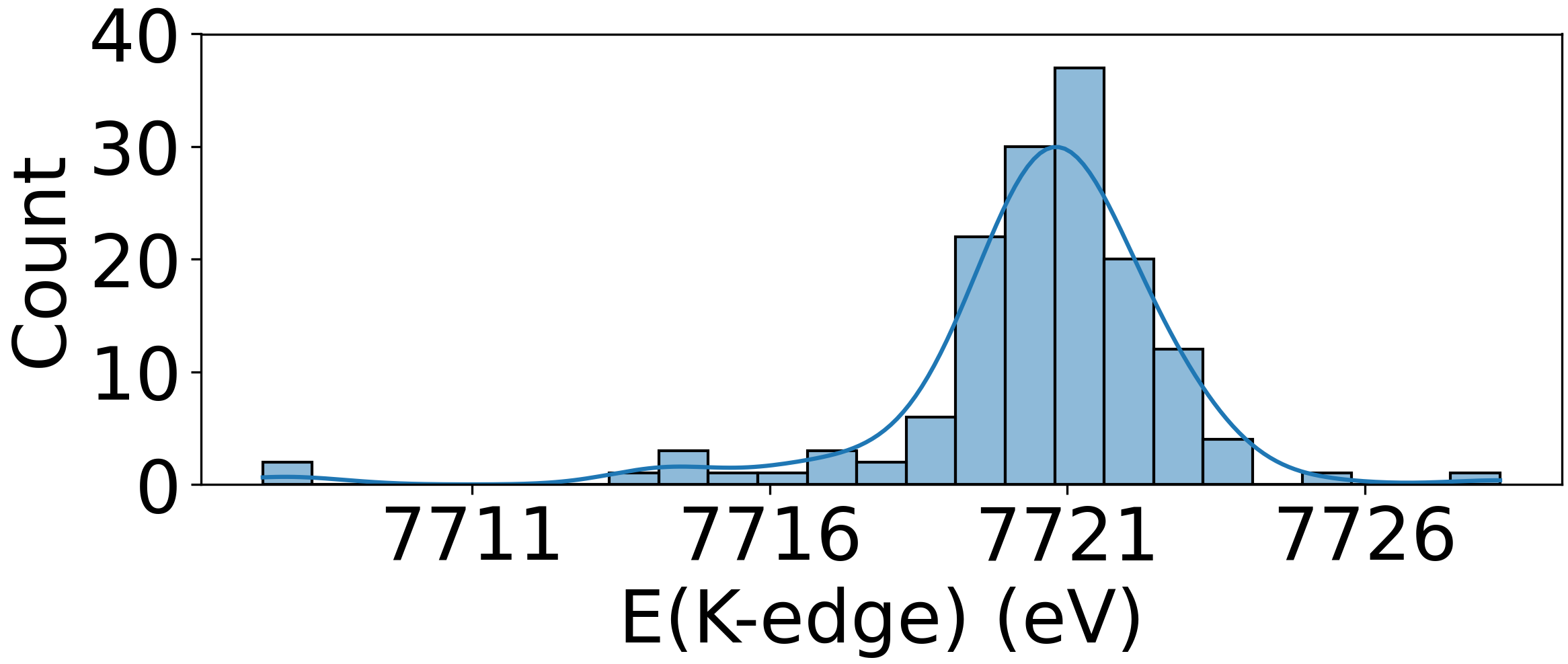}\\[0.5em]
        \subfigimg[width=\linewidth]{\textbf{(c)}}{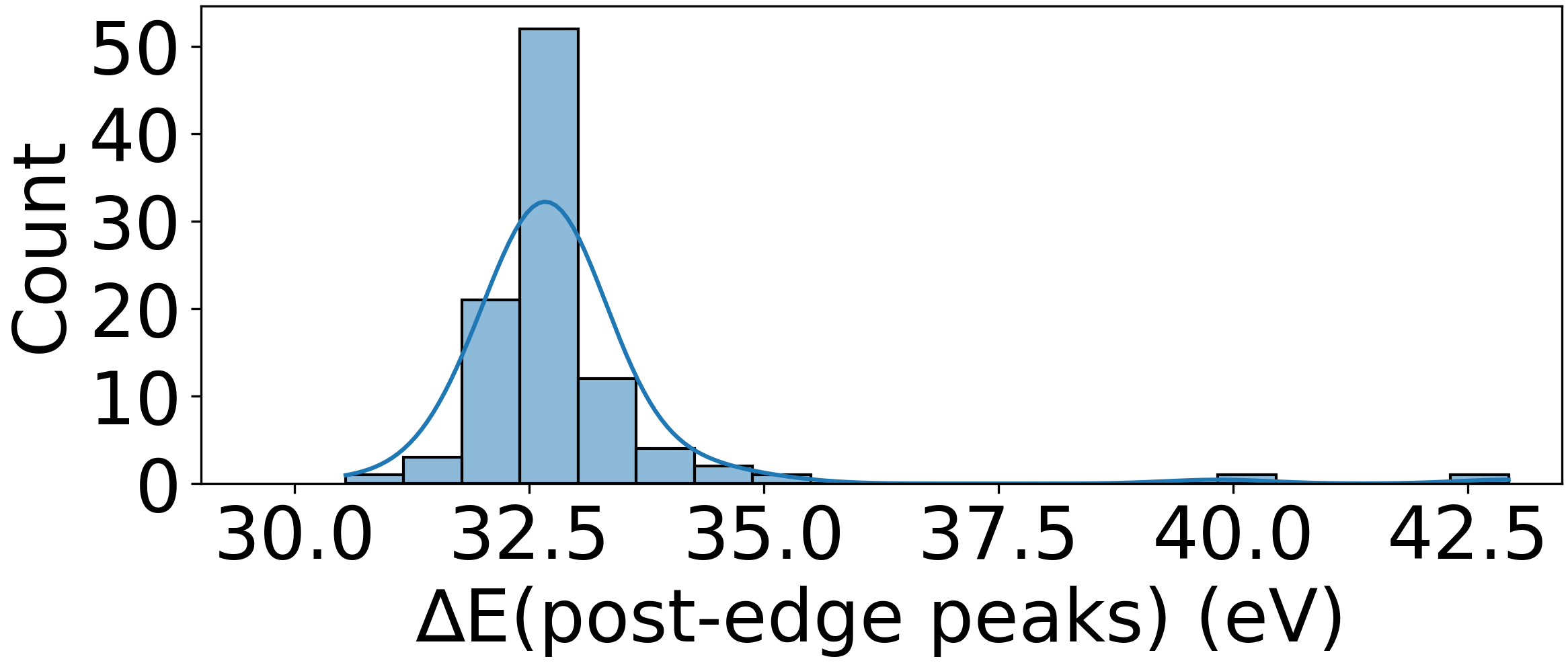}
    \end{minipage}
	\caption{
        Cross-laboratory comparison of 146 Co K-edge XAS spectra of CoO. 
        \textbf{(a)} Overlay of the 146 spectra, each vertically rescaled to set its highest peak to the same reference intensity. 
        \textbf{(b)} Histogram of edge energies, defined by the maximum first-derivative position in the edge region. 
        \textbf{(c)} Histogram of the energy spacing between the two adjacent post-edge peaks for spectra in which both peaks are resolved. 
	}
	\label{fig:XAS_CoO}
\end{figure}

Finally, the literature-mined dataset allows us to evaluate how measurements of the same material vary across laboratories, providing a valuable basis for standardization efforts.
As an example, we overlay 146 Co K-edge XAS spectra of CoO in Fig.~\ref{fig:XAS_CoO}(a).
Because XAS intensities are commonly reported in arbitrary units, each spectrum is vertically rescaled to set its highest peak to a common reference intensity, defined as the median peak intensity of comparable, consistently normalized literature spectra.
Inspection of the curves that deviate markedly from the typical pattern reveals a few anomalous spectra. 
For instance, one reports the Co(II) K-edge within 7100--7180~eV, likely a plotting error \cite{chen2023PrecisSolidp}, while another corresponds to CoO after electrochemical cycling and exhibits a post-edge structure more consistent with mixed Co$^{2+}$/Co$^{3+}$ states \cite{wu2021CobaltII}.
Spectra with clearly identifiable errors are excluded from Fig.~\ref{fig:XAS_CoO}, whereas the remaining outliers are retained. 
This example illustrates that aggregating spectra from many sources facilitates the identification of anomalies and outliers.

For each spectrum, the edge energy is defined as the energy at which the first derivative of the spectrum with respect to energy reaches its maximum within the edge region.
The distribution of edge energies (Fig.~\ref{fig:XAS_CoO}(b)) has a median of 7720.8~eV and an interquartile range (IQR; the spread between the 25th and 75th percentiles) \cite{tukey1977exploratory} of 1.8~eV.
The median and IQR are reported here because they are more robust to outliers than the mean (7720.5~eV) and standard deviation (2.4~eV).
In addition, we compute the energy spacing between two adjacent post-edge peaks (Fig.~\ref{fig:XAS_CoO}(c)).
This internal energy difference is less sensitive to absolute energy-calibration shifts while retaining structural relevance, as post-edge features primarily reflect multiple-scattering effects and the medium-range order around the absorber atom \cite{fdez-gubieda2016XrayAbsorption}.
The spacing has a median of 32.7~eV and an IQR of only 0.5~eV.
The smaller spread in this internal spacing suggests that the apparent variation in edge energy could be reduced through improved cross-laboratory alignment of the reference energy and corresponding shifts in the edge positions.
Because individual measurements are subject to fluctuations, statistics aggregated over many results provide a more reliable basis for identifying peaks and trends.
The dataset thus offers a quantitative reference for expected discrepancies across laboratories and informs the development of standards for cross-lab alignment and improved consistency.

Overall, we have constructed a large-scale, multimodal XAS dataset from the battery literature, pairing digitized spectra with essential measurement metadata as an AI-ready resource for foundation-model-scale analysis.
As a reference library, the dataset complements existing curated databases \cite{xaslib,ishii2025GlobalCrossd,spasyuk2025XASDBDesign,paripsa2024RefXASOpena,kieffer2016SSHADEFAME} with broader materials coverage for identifying trends in oxidation state, local coordination, and composition.
As training and evaluation data, the spectra and associated metadata can support machine-learning models for automated XAS interpretation and retrieval systems that connect new spectra to prior measurements.
Spanning thousands of source-traceable studies, the dataset also enables assessing experimental variability, harmonizing reporting practices, and linking spectroscopy to materials and their broader experimental context in multimodal knowledge bases.

\section*{Usage Notes}

The dataset is provided as a single JSONL file, readable with the standard libraries of major programming languages, including Python, MATLAB, R, and Wolfram Mathematica, with no additional dependencies.
For the spectral XY data, the values and units are preserved as reported in the original figure.
The energy unit of each record is given in its \texttt{x\_\allowbreak axis\_\allowbreak label} field; x values in keV can be converted to eV by multiplying by 1{,}000.
As absorption intensities are in arbitrary units, some figures omit numerical y-axis ticks; for these records, the y values are given in image coordinates, increasing from top to bottom of the image. 
When plotting such spectra, the y-axis can be inverted such that intensity increases upward, as conventionally displayed.
Each record can be traced back to its source paper and figure through the \texttt{doi} and \texttt{figure\_\allowbreak label} fields, if details beyond the dataset are needed.

\section*{Data Availability}

The dataset described in this Data Descriptor will be available at Figshare upon publication.

\section*{Code Availability}

The code used for XAS data digitization will be available on GitHub upon publication.

\bibliographystyle{naturemag}
\bibliography{refs}

\section*{Author Contributions}

Conceptualization: T.H., L.W., I.T.F., and M.K.Y.C.
Methodology: T.H. and A.V.
Investigation: T.H., A.V., and X.H.
Visualization: T.H. and Y.C.
Supervision: L.W., A.J., G.C., R.S.A., I.T.F., and M.K.Y.C.
Manuscript writing: T.H., I.T.F., and M.K.Y.C. drafted the manuscript; A.V., L.W., and A.J. provided critical revisions; all authors reviewed and contributed to the manuscript.

\section*{Competing Interests}

The authors declare no competing interests.

\section*{Acknowledgements}

We thank the Argonne Research Library for assistance in obtaining Text and Data Mining agreements with the specified publishers. 
We thank W. Jiang, K. Chard, R. Underwood, C. Siebenschuh, J. Huang, and H. Zheng for valuable discussions, and G. Burns III, T. Cohron, and A. Avarca for technical support with data management. 

\section*{Funding}

This work was supported by the U.S. Department of Energy, Office of Science, Office of Advanced Scientific Computing Research and Office of Basic Energy Sciences, Scientific Discovery through Advanced Computing (SciDAC) program under the FORUM-AI project. 
Work performed at the Center for Nanoscale Materials, a U.S. Department of Energy Office of Science User Facility, was supported by the U.S. DOE, Office of Basic Energy Sciences, under Contract No. DE-AC02-06CH11357.
Y.C. and M.K.Y.C acknowledge support by the Energy Storage Research Alliance "ESRA" (DE-AC02-06CH11357), an Energy Innovation Hub funded by the U.S. Department of Energy, Office of Science, Basic Energy Sciences.
T.H. acknowledges the support from Laboratory Directed Research and Development (LDRD) funding from Argonne National Laboratory, provided by the Director, Office of Science, of the U.S. Department of Energy under Contract No. DE-AC02-06CH11357.
This research used resources of the Argonne Leadership Computing Facility and the National Energy Research Scientific Computing Center, which are U.S. Department of Energy Office of Science User Facilities.

\end{document}